\documentclass{aa}
\usepackage{graphicx}
\usepackage{times}
\usepackage{txfonts}

\begin{document}
\title{The early-type close binary CV~Velorum 
revisited\thanks{Based on
    spectroscopic observations gathered with the CORALIE spectrograph mounted on
    the 1.2m Euler telescope at La Silla, Chile}}
\author{K. Yakut\inst{1,2},  C. Aerts\inst{1,3}
\and T. Morel\inst{1,}\thanks{External postdoctoral
      fellow of the European Space Agency}}
\offprints{conny@ster.kuleuven.be}

\institute{Instituut voor Sterrenkunde, Katholieke Universiteit
Leuven, Celestijnenlaan 200 B, B-3001 Leuven, Belgium \and
Department of Astronomy and Space Sciences, Ege University, 35100,
\.Izmir, Turkey \and Department of Astrophysics, University of
Nijmegen, PO Box 9010, 6500 GL Nijmegen, the Netherlands}

\date{Received .....; accepted .....}


\abstract{}{Our goal was to improve the fundamental parameters of the massive
  close double-lined eclipsing B2.5V+B2.5V binary CV~Velorum.}  {We gathered
  new high-resolution \'echelle spectroscopy on 13 { almost consecutive}
  nights covering essentially two orbits.  We computed a simultaneous solution
  to all the available { high-quality} radial-velocity and light data with
  the latest version of the Wilson-Deviney code.}  {We obtained the following
  values for the physical parameters: $M_1 = 6.066(74) M_\odot$, $M_2 =
  5.972(70) M_\odot$, $R_1 = 4.126(24) R_\odot$, $R_2 = 3.908(27) R_\odot$,
  $\log L_1 = 3.20(5) L_\odot$, and $\log L_2 = 3.14(5) L_\odot$. { The
  quoted errors contain a realistic estimate of systematic uncertainties mainly
  stemming from the effective temperature estimation. 
We derived abundances for both
  components and found them to be compatible with those of B stars in the solar
  neighbourhood.}  We discovered low-amplitude periodic line-profile variations
  with the orbital frequency for both components. Their interpretation requires
  new data with a longer time span.  The primary rotates subsynchronously while
  the secondary's $v\sin i$ and radius are compatible with synchronous
  rotation. Finally, we provide an update of the empirical mass-luminosity
  relation for main-sequence B stars which can be used for statistical
  predictions of masses or luminosities.  }{}  \maketitle

\section{Introduction}

Calibrating stellar evolution models requires to know the physical parameters of
stars with a very high accuracy.  Eclipsing and spectroscopic binaries with
detached components ensure to obtain the required precise parameters. In this
respect, CV~Vel (HD\,77464, Vmag = 6.7, spectral type B2.5V+B2.5V) and a few
other massive binary stars have an important place in astrophysics. The star CV
Vel is one of the well-known double-lined early-type eclipsing binary systems.

The binarity of CV~Vel was originally established by van Houten (1950). The
light curve of the system was first obtained by Gaposchkin (1955)
photographically, leading to the first estimate of the physical parameters of
the components.  The system was studied spectroscopically by Feast (1954) and
Andersen (1975). Feast (1954) obtained 34 spectra and derived the first
spectroscopic orbital parameters of the system. He reported the spectral type of
the components as B2V+B2V.  Andersen (1975) performed a spectral study of the
system based on data spread over 83 days and obtained much more accurate orbital
parameters. Clausen \& Gr{\o}nbech (1977) subsequently obtained a high quality
light curve of CV~Vel in the Str\"{o}mgren bandpasses. These authors derived the
orbital and physical parameters of the system adding the radial velocities of
Andersen (1975) to their data. The orbital period of the star amounts to
6.889494(8) days.

CV~Vel is an important massive binary whose parameters can { in principle\/}
be improved from modern high-quality data, which is why we revisited
this star after 30 years. We present a new detailed analysis of CV~Vel {
considering} all the available archival data to which we add new high-quality
\'{e}chelle spectra covering two consecutive weeks.  We also search for a
signature of oscillations by carefully analysing the line profiles of the
components, given that they reside in the instability strip of the slowly
pulsating B stars.

\section{New spectroscopic observations}
\begin{figure}
\centering
\rotatebox{0}{\resizebox{8.5cm}{!}{\includegraphics{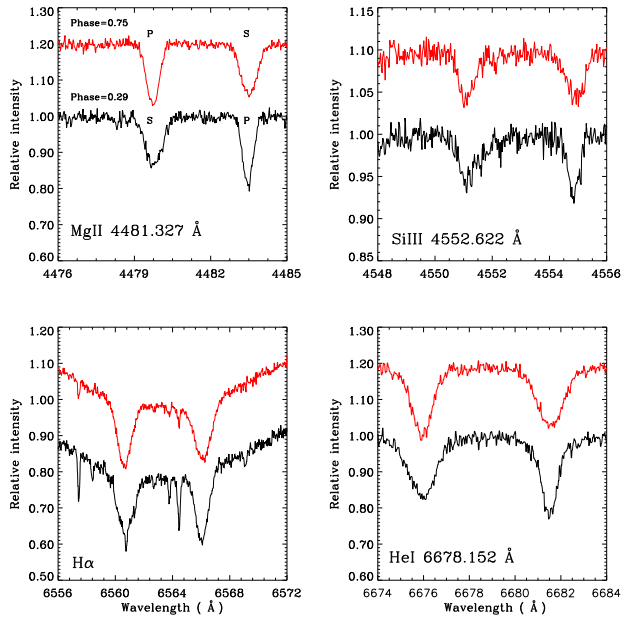}}}
\caption{Observed MgII 4481, SiIII 4552, H$\alpha$ and HeI 6678
line profiles at orbital phases 0.29 and 0.75. }
\label{fig1}
\end{figure}

New spectra of CV~Vel were obtained using the Swiss 1.2-m Euler telescope at La
Silla, Chile and the CORALIE spectrograph. The system was observed on 13 {
almost} successive nights in December 2001 and January 2002.  {
Table\,\ref{table1} lists the spectral lines used in our radial-velocity
analysis.} The resolution of the spectra is 50000 and the integration time was
in each case one hour to achieve an average S/N ratio near 200 in the blue part
of the spectrum.

The CORALIE spectra were reduced using the standard reduction pipeline (Baranne
et al., 1996). For the study of the orbital motion, Gaussian fits to the
absorption lines were made in order to derive their central wavelengths, thus
transforming into the radial velocities for each of the considered lines. 
To illustrate the data quality, we show in Fig.\,\ref{fig1} a
selection of four lines for the two orbital phases with the largest line
separation.  It is clear from this figure that the primary has narrower lines
than the secondary while being of similar temperature. This means that one of
them must be asynchronous { if the radii are similar,
as we show is the case below.}
We come back to this point in Sect.\,5.
\begin{table}
\caption{ Wavelength ranges and the spectral lines considered in our study.}
\begin{tabular}{llllll}
\end{tabular}
\begin{tabular}{lll}
$\lambda_{start}$ &  $\lambda_{end}$  &   considered lines   \\
\hline
4102 & 4148 &  HeI $\lambda$4121, HeI $\lambda$4126, SiII $\lambda$4128, SiII
$\lambda$4131, \\
&&HeI $\lambda$4144  \\
4464          &4514       &  HeI $\lambda$4471, MgII $\lambda$4481\\
4530          &4581       &
SiIII $\lambda$4553,SiIII $\lambda$4568,SiIII $\lambda$4575\\
      6528          &6600       &  H${\alpha}$\\
      6671          &6745       &  HeI $\lambda$6678\\
\hline
\end{tabular}
\label{table1}
\end{table}
\begin{table}
\caption{Radial velocity values for CV~Vel computed in this study.}
\begin{tabular}{llrrrr}
HJD(2452000+)&   Phase   &   V$_1$  &   O-C$_1$    &   V$_2$  &   O-C$_2$    \\
\hline \hline
272.6910    &   0.000   &           &       &   20.1    &   -3.7    \\
272.8130    &   0.017   &   -2.5?   &   -8.0    &   39.0    &   0.9 \\
273.6340    &   0.137   &   -76.9   &   -1.4    &   123.6   &   3.1 \\
273.8430    &   0.167   &   -87.2   &   2.0 &   135.2   &   0.9 \\
274.6680    &   0.287   &   -100.2  &   0.8 &   149.8   &   3.6 \\
274.8110    &   0.308   &   -95.4   &   0.5 &   142.3   &   1.2 \\
276.6810    &   0.579   &   83.0    &   -0.9    &   -39.1   &   2.4 \\
276.7990    &   0.596   &   95.8    &   0.4 &   -52.9   &   0.3 \\
277.5890    &   0.711   &   145.5   &   0.6 &   -103.1  &   0.4 \\
277.7320    &   0.731   &   148.4   &   0.7 &   -107.9  &   -1.7    \\
278.5650    &   0.852   &   120.5   &   -1.1    &   -78.0   &   1.8 \\
278.7740    &   0.883   &   104.8   &   -0.3    &   -65.4   &   -2.5    \\
279.5940    &   0.002   &           &           &   17.5?   &   -7.9    \\
279.7250    &   0.021   &   0.9     &   -2.0    &   39.2    &   -1.5    \\
279.8420    &   0.038   &   -8.4    &   1.9     &   54.0    &   -0.1    \\
280.5550    &   0.141   &   -78.9   &   -1.1    &   122.6   &   -0.2    \\
280.6970    &   0.162   &   -87.2   &   -0.1    &   132.8   &   0.6 \\
280.7620    &   0.171   &   -88.9   &   1.9 &   136.2   &   0.3 \\
281.5540    &   0.286   &   -100.7  &   0.4 &   146.3   &   0.0 \\
281.6820    &   0.305   &   -96.6   &   0.1 &   140.0   &   -1.9    \\
281.7840    &   0.320   &   -91.5   &   0.5 &   134.4   &   -2.7    \\
282.5610    &   0.433   &   -26.7   &   1.6 &   73.7    &   1.2 \\
282.6970    &   0.452   &   -12.8   &   0.9 &   59.3    &   1.7 \\
283.5830    &   0.581   &   85.1    &   -0.1    &   -46.8   &   -4.0    \\
283.6840    &   0.596   &   95.3    &   0.3 &   -54.4   &   -1.6    \\
283.8000    &   0.612   &   105.9   &   0.4 &   -62.9   &   0.5 \\
284.5480    &   0.721   &   147.3   &   0.8 &   -103.6  &   1.5 \\
284.6990    &   0.743   &   148.8   &   0.6 &   -103.2  &   3.6 \\
284.7720    &   0.753   &   149.0   &   0.8 &   -103.0  &   3.8 \\
285.6550    &   0.882   &   107.4   &   1.6 &   -65.7   &   -2.0    \\
\hline
\end{tabular}
\label{table2}
\end{table}

We list in Table\,\ref{table2} the average value of the measured radial
velocities obtained from the five least blended spectral lines, i.e.\ He I
4120.993\AA, Mg II 4481.327\AA, Si III 4552.622\AA, Si III 4567.840\AA\ and He I
6678.149\AA\ lines. For each of these lines, the orbit was solved separately and
the obtained results are summarized in Table\,\ref{table3}. { The 
internal precision}
of a single observation amounts to $\sim$ 1.0 km\,s$^{-1}$ for the primary and
$\sim$ 2.0\,km\,s$^{-1}$ for the secondary. We note a quite significant
deviation among the $V_0$ values derived from the different lines. This is not
uncommon in this part of the HR diagram (e.g.\ Mathias et al.\ (1994) for a
discussion). It is interpreted in various ways, among which unknown blends
within the lines, an asymmetric Stark effect among different lines (Struve \&
Zebergs 1960) or the presence of a so-called {\it Van Hoof effect\/} (Van Hoof
1975) in pulsating stars. The latter effect is due to the fact that spectral
lines form in different line forming regions in the stellar atmosphere, such
that a running wave nature of the oscillations implies a shift in
phase during the velocity cycle. Such an effect has clearly been detected in the
pulsating B stars BW\,Vulpeculae and $\alpha\,$Lupi (Mathias \& Gillet
1993). Our data have insufficient time base and temporal resolution to test the
presence of a Van Hoof effect in CV~Vel (see also Section~6).  Even if a Van
Hoof effect is present, it would not explain the difference in $V_0$ for the two
Si lines of the same triplet (SiIII $\lambda$4553 and SiIII $\lambda$4568).  A
similar unexplained average velocity shift between the same lines in this Si
triplet was encountered by De Cat et al.\ (2000) in their analysis of several
hundred line-profile data, of very high S/N, of the B0.5III star
$\varepsilon\,$Per. As there, we are unable to offer a good interpretation of
this shift.  We will take it into account as a systematic uncertainty in the
derivation of the orbital and fundamental parameters of the star (see below).  
Though
the eccentricity, $e$, was first regarded as a free parameter, its value always
appeared to be zero, within the error range, so we do not list it in
Table\,\ref{table3}.

\begin{figure}
\centering
\rotatebox{0}{\resizebox{8.5cm}{!}{\includegraphics[bb=90 360 400 650]{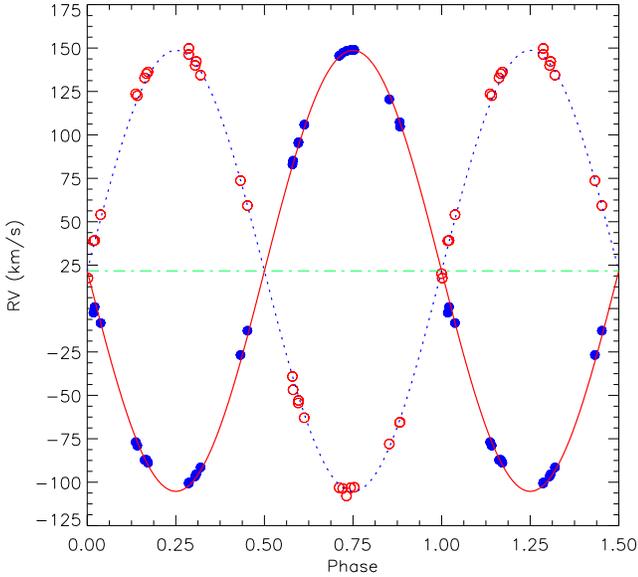}}}
\caption{The radial velocity observations of CV~Vel versus the
orbital phase. The filled and open circles correspond to the
velocities of the primary and the secondary, respectively. The data
are listed in Table\,\protect\ref{table2} and the sine curve 
corresponds to the elements given in Table\,\protect\ref{table4}.
The center-of-mass velocity is indicated by the { dashed-dotted}
line.}\label{fig2}
\end{figure}

The { fifth} column of Table\,\ref{table4} gives the orbital solution from
the averaged radial velocities of our study. The average radial-velocity values
are plotted versus the orbital phase for each of the components in
Fig.\,\ref{fig2}. The phases were calculated using the ephemeris given by
Clausen \& Gr{\o}nbech (1977):
\begin{equation}
\textrm{Pri. Min.} = \textrm{HJD} \,2442048\fd66894(14) +
6\fd889494(8)\times E.
\label{eph1}
\end{equation}

\begin{table*} \caption{Spectroscopic orbital elements of CV~Vel for different
spectral lines. See text for details. The formal standard errors $\sigma$ are
given in parentheses in the last digit quoted.}
\begin{tabular}{lllllllll}
\hline
Ion     & Lines      &  K$_1$     &  K$_2$        &m$_1$sin$^3i$   &
m$_2$sin$^3i$  & a$_1$sin$i$     & a$_2$sin$i$  & $V_0$     \\
        &\AA   & km\,s$^{-1}$ & km\,s$^{-1}$         &$\rm{M_{\odot}}$&
$\rm{M_{\odot}}$& $\rm{R_{\odot}}$& $\rm{R_{\odot}}$ &   km\,s$^{-1}$ \\
\hline
He I    &4120.993    & 126.8(5)   & 128.6(5)      &5.988           & 5.904
&17.261           &17.545   &  16.2(2)         \\
Mg II   &4481.327    & 126.9(4)   & 128.6(4)      &5.993           & 5.913
&17.290           &17.528    &  18.2(2)        \\
Si III  &4552.622    & 125.2(6)   & 128.2(6)      &5.876           & 5.739
&17.037           &17.447 &  23.5(3)           \\
Si III  &4567.840    & 127.0(5)   & 128.2(5)      &5.960           & 5.904
&17.315           &17.473   &  24.7(3)         \\
He I    &6678.149    & 127.0(4)   & 128.6(4)      &5.997           & 5.923
&17.287           &17.498    &  24.2(2)        \\
\hline
\end{tabular}
\label{table3}
\end{table*}

\begin{table*}
\caption{Spectroscopic orbital parameters of CV~Vel.  The standard
errors $\sigma$ are given in parentheses in the last digit quoted.}
\tabcolsep=4pt
\begin{tabular}{lllllll}
\hline
Parameter & Unit & Feast (1954) & Andersen (1975) & This paper & 
This paper $+$ Andersen (1975) \\
\hline
K$_1$ & km\,s$^{-1}$ & 122.2(6)  & 127.0(3) & 126.5(3) & 127.0(2) \\
K$_2$ & km\,s$^{-1}$ & 126.6(6)  & 129.2(4) & 128.6(3) & 129.1(2) \\
V$_0$ & km\,s$^{-1}$ & 27.9(1.1) & 24.3(5)  & 21.7(2)  &  21.7(1) \\
q &${m_2}/{m_1}$   & 0.965     & 0.983    & 0.984    &  0.984   \\
a$\sin$$i$  & $\rm{R_{\odot}}$ & 33.72    & 34.87    & 34.722 &  34.857 \\
m$_1$sin$^3i$ & $\rm{M_{\odot}}$ & 5.61   & 6.07     & 5.974  &  6.021 \\
m$_2$sin$^3i$ & $\rm{M_{\odot}}$ & 5.41   & 5.96     & 5.876  &  5.927 \\
$\sum (O-C)^2 $  &  & 0.08 & 0.01   & 0.01     &  0.01     \\
\hline
\end{tabular}
\label{table4}
\end{table*}

\section{Simultaneous radial-velocity and light curve analysis}

{ To measure the $\sum \omega (O-C)^2 $ values, given in Table\,\ref{table4},
we have re-analyzed the data of Feast (1954) and Andersen (1975). This showed
that the data of Andersen (1975) and those of our paper have similar quality
while those of Feast (1954) have clearly lower quality, as already emphasised by
Andersen (1975).  For this reason, we omitted Feast's data in the remaining of
this paper.

We combined the newly obtained radial-velocity values (Table\,2) with those
previously obtained by Anderson (1975). This gave a total of 62 radial-velocity
values for each of the components to which we assigned equal weights. As can be
seen in Table\,\ref{table4}, we find an offset in average velocity $V_0$ between
the two radial-velocity sets. This could be due to the difference in used
spectral lines in the two studies, but it may also be that we find a downward
trend in $V_0$ over time due to the presence of a third body. We have too few
epochs to model any $V_0$ trend for the time being. For this reason, we shifted
the data towards the CORALIE $V_0$ value, keeping in mind that this systematic
uncertainty adds to the uncertainty of the overall solution.}

These 62 radial-velocity values were then combined with the light curves
obtained in the Str\"{o}mgren (ubvy) filters by Clausen \& Gr{\o}nbech (1977),
kindly made available to us by these authors.  
{ The u band data show a
somewhat larger scatter than the other filters so we used different weights. We
determined the weights for the uvby bands from the observing errors given by
Clausen \& Gr{\o}nbech (1977), resulting in the values 1.7, 2.5, 2.0, and 2.0
for u, v, b and y, respectively.}

The previous light curve analyses of the system have been done using the WINK
method by Clausen \& Gr{\o}nbech (1977) (second column of Table\,5). 
Clausen \& Gr{\o}nbech (1977) have used a linear limb-darkening law, fixing the
temperature of the primary star at the value 18300~K. They determined solutions
for each colour, assuming different values for $k=\frac{r_2}{r_1}$ (see, their
Table~4).  In this work, the light and radial-velocity curves were analyzed
simultaneously using the 2005 September version of the Wilson-Deviney code
(hereafter WD, Wilson \& Devinney 1971; Wilson 1994). Mode 2 of the WD code,
which assumes that both of the components are detached, was adopted. In the
final solution, we fixed the following parameters: the limb-darkening
coefficients { (taken from Diaz-Cordoves \& Gimenez 1992, see Table\,5)}, the
values of the gravity darkening coefficients and the albedos. The adjustable
parameters are the inclination $i$, the temperature of the secondary component
$T_{2}$, the luminosities $L_{1b}$ and $L_{1y}$, and the surface potentials
$\Omega_1$ and $\Omega_2$. The differential correction (DC) code was ran until
the corrections to the input parameters were lower than their errors.

\begin{figure}
\centering
\rotatebox{0}{\resizebox{8.5cm}{!}{\includegraphics[bb=80 380 340 850]{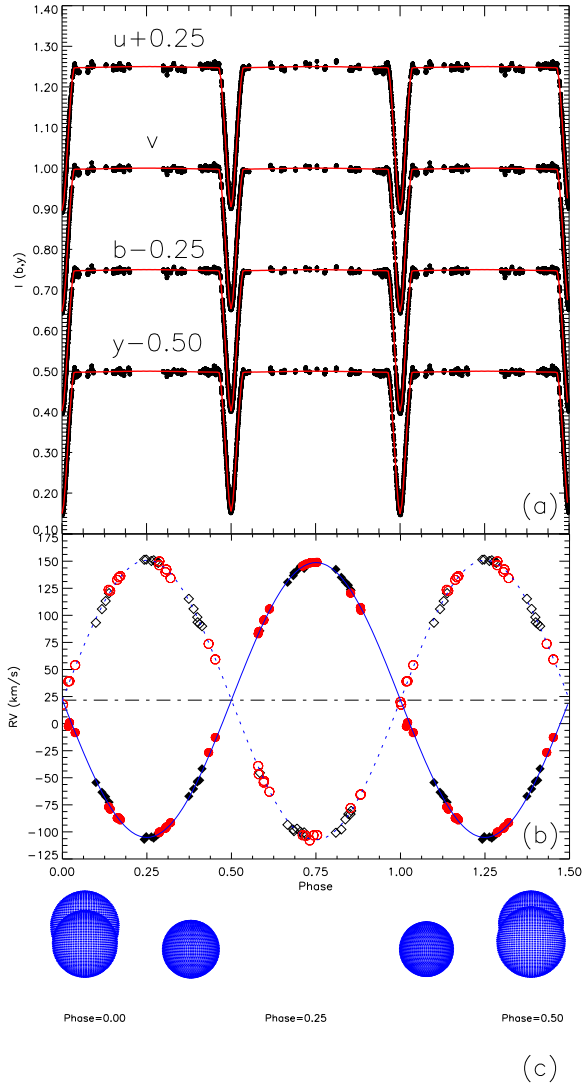}}}\\[-.50cm]
\caption{The observed and computed (a) light and (b) radial-velocity
curves of CV~Vel.  The curve in the u, b and y bands are moved
by +0.25, -0.25 and -0.50, respectively, in intensity for good
visibility. Circles and squares denote the data in this study, and
Anderson (1975) respectively.
{ The center-of-mass velocity is indicated by the dashed-dotted
line.} (c) 3-D representation of CV~Vel at
phases 0.00, 0.25 and 0.50.}
\label{fig3}
\end{figure}

{ To obtain a simultaneous solution, different possibilities were tested. We
varied the primary's temperature in the range of 17800 K - 19000~K with 50~K
steps and found that T$_1$=18000~K gave the best solution with the lowest formal
errors for the unknown quantities and the lowest $\sum (O-C)^2$. However, the
solutions with $T_1$ between 18000~K and 19000~K were almost indistinghuishable.
This is in agreement with the solution proposed by Clausen \& Gr{\o}nbech
(1977), who assigned $T_1=18200$~K to the { mean component of the system}
and considered an uncertainty
of 500~K. In selecting the best solution, we first assumed the system to be
eccentric leaving the $e$ and $\omega$ values free. As a result we obtained
always $e \leq 0.001$.  The final results are given in Table\,\ref{table5} for
each band separately, as well as for the simultaneous solution including all
photometric data.  We find a similar orbital inclination value than Clausen \&
Gr{\o}nbech (1977).}
The computed light and radial-velocity curves for the parameters given in
Table\,\ref{table5} are shown by the solid lines, and are compared with all the
observations, in Fig.\,\ref{fig3}.

\begin{table*}
\caption{ Orbital parameters and photometric elements, with their formal
  errors,  of CV~Vel. See text for
details. CG: Clausen \& Gr{\o}nbech (1977), the other columns represent results
obtained in this work. The parameters with a superscript $\ast$ were kept
fixed.}
\begin{tabular}{lllllll}
\hline
 &  CG             & u                  & v              & b             & y              & ubvy                \\
Parameter                           & Value ( $\sigma$ )& Value ($\sigma$ )   & Value ( $\sigma$ )& Value ($\sigma$ ) & Value ( $\sigma$ )& Value ($\sigma$ ) \\
\hline
Geometric parameters:                &                 &                    &                &               &                &                      \\
$i$ ${({^\circ})}$                   &   86.59(5)      &    86.66(2)        &    86.62(2)    &    86.62(2)   &    86.62(2)    &    86.63(2)          \\
$\Omega _{1}$                        &    -            &    9.500(9)        &    9.445(9)    &    9.451(9)   &    9.445(9)    &    9.456(10)          \\
$\Omega _{2}$                        &    -            &    9.852(11)       &    9.807(11)   &    9.805(11)  &    9.794(11)   &    9.804(12)         \\
$a$                                  &    -            &    34.897(19)      &    34.896(19)  &    34.891(19) &    34.898(19)  &    34.899(20)        \\
$q$                                  &    -            &    0.9837(9)       &    0.9842(9)   &    0.9845(9)  &    0.9838(9)   &    0.9835(10)         \\
Fractional radii of the primary      &                 &                    &                &               &                &                      \\
${r}_{1~pole}$                       &   -             &   0.1172(4)        &   0.1180(3)    &   0.1181(3)   &   0.1182(3)    &   0.1172(1)          \\
${r}_{1~point}$                      &   -             &   0.1177(4)        &   0.1186(3)    &   0.1186(3)   &   0.1187(3)    &   0.1184(1)          \\
${r}_{1~side}$                       &   -             &   0.1174(4)        &   0.1182(3)    &   0.1183(3)   &   0.1184(3)    &   0.1181(1)          \\
${r}_{1~back}$                       &   -             &   0.1176(4)        &   0.1185(3)    &   0.1186(3)   &   0.1186(3)    &   0.1184(1)          \\
${r}_{1~mean}$                       &   0.117(1)      &   0.1175(4)        &   0.1183(3)    &   0.1184(3)   &   0.1185(3)    &   0.1182(1)          \\
Fractional radii of the secondary    &                 &                    &                &               &                &                      \\
${r}_{2~pole}$                       &   -             &   0.1113(4)        &   0.1117(3)    &   0.1118(3)   &   0.1118(3)    &   0.1117(2)          \\
${r}_{2~point}$                      &   -             &   0.1117(4)        &   0.1122(3)    &   0.1122(3)   &   0.1123(3)    &   0.1121(2)          \\
${r}_{2~side}$                       &   -             &   0.1115(4)        &   0.1119(3)    &   0.1119(3)   &   0.1120(3)    &   0.1119(2)          \\
${r}_{2~back}$                       &   -             &   0.1117(4)        &   0.1121(3)    &   0.1121(3)   &   0.1122(3)    &   0.1121(2)          \\
${r}_{2~mean}$                       &   0.113(1)      &   0.1115(4)        &   0.1120(3)    &   0.1120(3)   &   0.1121(3)    &   0.1119(2)          \\
Radiative parameters:                &                 &                    &                &               &                &                      \\
$T_1$$^*$ (K)                        &   18200   &   18000            &   18000        &   18000       &   18000        &   18000              \\
$T_2$ (K)                            &   18060         &   17813(50)        &   17703(50)    &   17818(50)   &   17815(50)    &   17790(50)          \\
Albedo$^*$ ($A_1=A_2$)               &   1.0           &   1.0              &   1.0          &   1.0         &   1.0          &   1.0                 \\
Limb darkening coefficients$^*$:     &                 &                    &                &               &                &                       \\
Square Law:                          &                 &                    &                &               &                &                        \\
$x_{1u}^{bol}$                       & -               &    0.054           &                &               &                &    0.054               \\
$x_{1v}^{bol}$                       & -               &                    &   -0.089       &               &                &   -0.089               \\
$x_{1b}^{bol}$                       & -               &                    &                &   -0.078      &                &   -0.078               \\
$x_{1y}^{bol}$                       & -               &                    &                &               &   -0.067       &   -0.067               \\
$x_{2u}^{bol}$                       & -               &    0.059           &                &               &                &    0.059               \\
$x_{2v}^{bol}$                       & -               &                    &   -0.073       &               &                &   -0.073               \\
$x_{2b}^{bol}$                       & -               &                    &                &   -0.078      &                &   -0.078               \\
$x_{2y}^{bol}$                       & -               &                    &                &               &   -0.067       &   -0.067               \\
$y_{1u}^{bol}$                       & -               &   0.492            &                &               &                &   0.492               \\
$y_{1v}^{bol}$                       & -               &                    &   0.718        &               &                &   0.718                \\
$y_{1b}^{bol}$                       & -               &                    &                &   0.666       &                &   0.666                \\
$y_{1y}^{bol}$                       & -               &                    &                &               &   0.581        &   0.581                \\
$y_{2u}^{bol}$                       & -               &   0.550            &                &               &                &   0.550                \\
$y_{2v}^{bol}$                       & -               &                    &   0.611        &               &                &   0.611                \\
$y_{2b}^{bol}$                       & -               &                    &                &   0.663       &                &   0.663                \\
$y_{2y}^{bol}$                       & -               &                    &                &               &   0.583        &   0.583                \\
Gravity brightening$^*$ ($g_1=g_2$)  & 1.0             &   1.0              &   1.0          &   1.0         &   1.0          &   1.0                  \\
Luminosity ratio                     &                 &                    &                &               &                &                        \\
$(\frac{L_1}{L_1 +L_2})_u$           & 0.519(20)       &   0.533(13)        &                &               &                &   0.534(13)            \\
$(\frac{L_1}{L_1 +L_2})_v$           & 0.517(20)       &                    &   0.531(12)    &               &                &   0.528(12)            \\
$(\frac{L_1}{L_1 +L_2})_b$           & 0.517(20)       &                    &                &   0.532(12)   &                &   0.532(12)            \\
$(\frac{L_1}{L_1 +L_2})_y$           & 0.516(20)       &                    &                &               &   0.531(12)    &   0.532(12)             \\
$\sum (O-C)^2  $                     & -               &   0.028            &   0.026        &   0.023       &   0.025        &   0.064                \\
 \hline
\end{tabular}
\label{table5}
\end{table*}

\section{Abundance determination}

{ 

Our spectra are of sufficient quality, and cover a wide enough wavelength range,
to undertake an abundance analysis for the two components of the CV~Vel system.
The non-local thermodynamic equilibrium (NLTE) abundances of He, C, N, O, Mg,
Al, Si, S and Fe were calculated using the latest versions of the line formation
codes DETAIL/SURFACE and plane-parallel, fully line-blanketed Kurucz atmospheric
models (Kurucz \cite{kurucz93}). Curve-of-growth techniques were used to
determine the abundances using the equivalent widths (EWs) of a set of spectral
lines measured on a mean spectrum created by averaging the five CORALIE spectra
taken at $\phi$$\sim$0.73 (see Morel et al. \cite{morel} for the line list
used). These EWs were subsequently corrected to account for the dilution by the
companion's continuum, assuming the luminosity ratio given in
Table\,\ref{table7}. The two components have essentially the same effective
temperature, so that these correcting factors can be considered independent of
wavelength.

We first computed the abundances assuming the $\log g$ and $T_{\rm eff}$ values
obtained from the light curve analysis (Table\,\ref{table7}), but found that the
\ion{Si}{ii/iii} ionization balance is not well
fullfilled in that case. Instead, a
slightly higher value of $T_{\rm eff}$=19\,000 $\pm$ 500 K is needed for both
components. Because of this ambiguity surrounding the temperature determination,
we have estimated the abundances using both the photometric and the
spectroscopic values (hereafter $T_{\rm photo}$ and $T_{\rm ionization}$,
respectively). The uncertainties on the elemental abundances were calculated by
adding in quadrature the internal errors (i.e.\ the line-to-line scatter) and
the errors arising from the uncertainty on the microturbulent velocity,
$\xi$. The latter quantity is usually inferred in B stars by requiring no
dependence between the abundances yielded by lines of an ionic species with
strong features (e.g.\ \ion{O}{ii}) and their strength. However, most metallic
lines are intrinsically weak in the relevant temperature range (measured EWs
$\lesssim$ 20 m\AA), such that it is virtually impossible to constrain the
microturbulence. We adopt in the following a canonical value $\xi$=5$\pm$5 km
s$^{-1}$, with the large conservative error bars encompassing all plausible
values for B dwarfs. This translates into poorly defined abundances for the
elements solely exhibiting relatively strong lines, in particular He, Si and Mg.

The results for the two choices of $T_{\rm eff}$ are provided in
Table~\ref{tab_abundances}. No abundances are found to deviate at more than the
3$\sigma$ level from the typical values found for early B dwarfs in the solar
neighbourhood (Daflon \& Cunha \cite{daflon_cunha}) for {\it both} $T_{\rm eff}$
estimates. Therefore, there is no convincing evidence for departures from a
scaled solar composition. The nominal abundance values are systematically higher
in the secondary than in the primary. As the chemical composition of the two
components is expected to be identical for such a young, detached system with no
past or ongoing episodes of mass transfer, this strongly argues for a higher
microturbulent velocity in the secondary. This difference is, however, not
sufficient to explain the discrepant rotation rates of the two components
(see Section\,6 below). 

Recent works point to the existence of a population of
slowly-rotating B-type (magnetic) pulsators exhibiting an unexpected nitrogen
excess at their surface, which could possibly be attributed to deep mixing or
diffusion effects (Briquet \& Morel \cite{briquet}; Morel et
al.\ \cite{morel}). A similar phenomenon cannot be firmly established in either
of the two components of CV~Vel: while the ratio of the abundances of N and C
([N/C]) is comparable to the values previously reported for the N-rich $\beta$
Cephei and slowly pulsating B stars, the [N/O] ratio is indistinguishable from
solar. This conclusion holds irrespective of the temperature scale adopted.

\begin{table*}
\centering
\caption{Mean NLTE abundances of the primary and secondary components of CV~Vel
(on the scale in which $\log \epsilon$[H]=12), along with the total 1-$\sigma$
uncertainties. The number of used lines is given in brackets. We quote the
values obtained when using the temperatures derived from the light curve
analysis ($T_{\rm photo}$) or from the ionization balance of silicon ($T_{\rm
ionization}$). The gravity is fixed in both cases to the values quoted in
Table\,\protect\ref{table7} and the microturbulent velocity, $\xi$, to 5 km
s$^{-1}$. For comparison purposes, the first column gives the typical values
found for nearby B dwarfs (Daflon \& Cunha \cite{daflon_cunha}). We define [N/C]
and [N/O] as $\log$[$\epsilon$(N)/$\epsilon$(C)] and
$\log$[$\epsilon$(N)/$\epsilon$(O)], respectively.}
\label{tab_abundances}
\begin{tabular}{lccccc} \hline\hline
&  & \multicolumn{2}{c}{$T_{\rm eff}$$\equiv$$T_{\rm photo}$=18\,000 K}   & \multicolumn{2}{c}{$T_{\rm eff}$$\equiv$$T_{\rm ionization}$=19\,000 K}\\
                    &  OB stars      & Primary             & Secondary           & Primary             & Secondary \\\hline
He/H                &  0.085         & 0.053$\pm$0.029 (6) & 0.097$\pm$0.033 (6) & 0.040$\pm$0.022 (6) & 0.073$\pm$0.031 (6)\\
$\log \epsilon$(C)  &  $\sim$8.2     & 8.08$\pm$0.10 (4)   & 8.29$\pm$0.17 (2)   & 7.89$\pm$0.07 (4)   & 8.11$\pm$0.17 (2)\\
$\log \epsilon$(N)  &  $\sim$7.6     & 7.93$\pm$0.13 (7)   & 8.16$\pm$0.16 (6)   & 7.69$\pm$0.13 (7)   & 7.92$\pm$0.17 (6)\\
$\log \epsilon$(O)  &  $\sim$8.5     & 8.74$\pm$0.18 (8)   & 8.85$\pm$0.20 (4)   & 8.44$\pm$0.17 (8)   & 8.55$\pm$0.19 (4)\\
$\log \epsilon$(Mg) &  $\sim$7.4     & 7.01$\pm$0.40 (1)   & 7.26$\pm$0.45 (1)   & 7.13$\pm$0.40 (1)   & 7.39$\pm$0.39 (1)\\
$\log \epsilon$(Al) &  $\sim$6.1     & 6.32$\pm$0.18 (3)   & 6.48$\pm$0.17 (3)   & 6.12$\pm$0.18 (3)   & 6.27$\pm$0.17 (3)\\
$\log \epsilon$(Si) &  $\sim$7.2     & 6.97$\pm$0.44 (7)   & 7.15$\pm$0.61 (7)   & 6.93$\pm$0.28 (7)   & 7.11$\pm$0.44 (7)\\
$\log \epsilon$(S)  &  $\sim$7.2     & 7.02$\pm$0.21 (5)   & 7.25$\pm$0.23 (2)   & 7.08$\pm$0.19 (5)   & 7.29$\pm$0.25 (2)\\
$\log \epsilon$(Fe) &  $\sim$7.3$^a$ & 7.20$\pm$0.20 (3)   & 7.46$\pm$0.18 (2)   & 7.05$\pm$0.15 (3)   & 7.34$\pm$0.23 (2)\\\hline
${\rm [N/C]}$       &  $\sim$--0.6   & --0.15$\pm$0.17     & --0.13$\pm$0.24     & --0.20$\pm$0.15     & --0.19$\pm$0.25 \\
${\rm [N/O]}$       &  $\sim$--0.9   & --0.81$\pm$0.23     & --0.69$\pm$0.26     & --0.75$\pm$0.22     & --0.63$\pm$0.26 \\
\hline
\end{tabular}
\begin{flushleft}
$^a$ From Morel et al. (\cite{morel}).
\end{flushleft}
\end{table*}

Regarding the derivation of the physical parameters of the component stars, we
are thus left with the conclusion that a second systematic uncertainty occurs,
besides the one induced by the offset in $V_0$. We forced the effective
temperature of the primary to 19\,000 K and checked the orbital solutions from
WD discussed in the previous section. As already mentioned, the differences in
the orbital parameter values are small and compatible with the uncertainties
listed there. While deriving the physical parameters of the system in the next
section, we propagated the errors caused by the systematic uncertainty in
effective temperature in order to achieve realistic errors.  }

\section{Physical parameters of the system}

The physical parameters of CV~Vel first obtained by Gaposchkin (1955) were
$M_1=5.618\,M_\odot$, $M_2=5.418\,M_\odot$, $R_1=4.26\,R_\odot$, and
$R_2=4.16\,R_\odot$. Andersen (1975) reported the parameters of the components
as $M_{1,2} = 6.05\,M_\odot$, $R_{1,2} = 4.05\,R_\odot$, and $\log\,T_{\rm
eff}=4.26$. Clausen \& Gr{\o}nbech (1977) subsequently obtained
$M_1=6.10\,M_\odot$, $M_2=5.99\,M_\odot$, $R_1=4.10\,R_\odot$,
$R_2=3.95\,R_\odot$, $T_{1} = 18200$\,K and $T_{2} = 18060$\,K.

We summarize the obtained parameters of the system and of the components
obtained from our simultaneous fit in Table\,\ref{table7}.  { In computing
them, we took the values of the light curve estimates for the effective
temperature of the components. We argue that the photometric data are of high
quality while the
spectroscopic temperature estimates suffer from the unknown microturbulence.
However, we take into account a systematic uncertainty of 500\,K for the
temperature. This equals the uncertainty derived from the silicon ionisation
balance and is a factor ten higher than the formal error that resulted from our
WD analysis, so we consider this to be a safe conservative approach.  We
propagated this uncertainty for all the derived quantities relying on the
effective temperature.}  

We derived the component's bolometric corrections from the temperatures using
the interpolation formula given by Balona (1994) for G5 to early-type stars.  In
order to find the distance of the system, we used the E(B-V) value of $0\fm08$
from Savage et al.\ (1985) resulting in $d=553$\,pc.

{ 
Our results are in very good agreement with
those obtained by Clausen \& Gr\o nbech (1977).
We were unable to 
achieve more precise values than they did, due to the 
uncertainty on the effective temperature resulting from our spectroscopic
analysis. }

\begin{table*}
\caption{ Absolute parameters for CV~Vel. The errors are given in parentheses
and take into account the systematic uncertainties encountered for $V_0$ and for
the effective temperature of the components.  The effective temperature of the
Sun was taken as 5780~K, its bolometric absolute magnitude as $4.75$ mag and its
bolometric correction as --0.07 mag.  }
\label{table7}
\begin{tabular}{llll}
\hline
 &Unit                  & primary component      & secondary component     \\
\hline
Mass ($M$)                                        &$\rm{M_{\odot}}$      & $6.066\,(74)$          & $5.972\,(70)$     \\
Radius ($R$)                                      &$\rm{R_{\odot}}$      & $4.126\,(24)$          & $3.908\,(27)$    \\
Effective temperature ($\log T_{\rm eff}$)        &$\rm{K}$              & $4.255\,(12)$          & $4.250\,(12)$     \\
Luminosity ($\log L$)                             &$\rm{L_{\odot}} $     & $3.204\,(48)$          & $3.137\,(48)$      \\
Surface gravity $(log\,g)$                        &cgs                   &  3.99\,(6)             &  4.03\,(6)         \\
Bolometric correction (BC)                        &mag                   & -1.68\,(5)           & -1.66\,(5)         \\
Absolute bolometric magnitude (M$_{bol}$)         &mag                   & -3.26\,(3)             & -3.09\,(3)         \\
Absolute visual magnitude (M$_{V}$)               &mag                   & -1.58\,(6)             & -1.43\,(5)           \\
Semi$-$major axis ($a$)                           &$\rm{R_{\odot}}$      &\,\,\,\,\,\,\,\,\,\,\,\,\,\,\,\,\,\,\,\,\,\,\,\,\,\,34.90\,(15) &   \\
Distance ($d$)                                    &pc                    &\,\,\,\,\,\,\,\,\,\,\,\,\,\,\,\,\,\,\,\,\,\,\,\,\,\,\,\,553\,(32)    &   \\
\hline
\end{tabular}
\label{table6}
\end{table*}

\begin{figure}
\centering
\rotatebox{-90}{\resizebox{10.5cm}{!}{\includegraphics{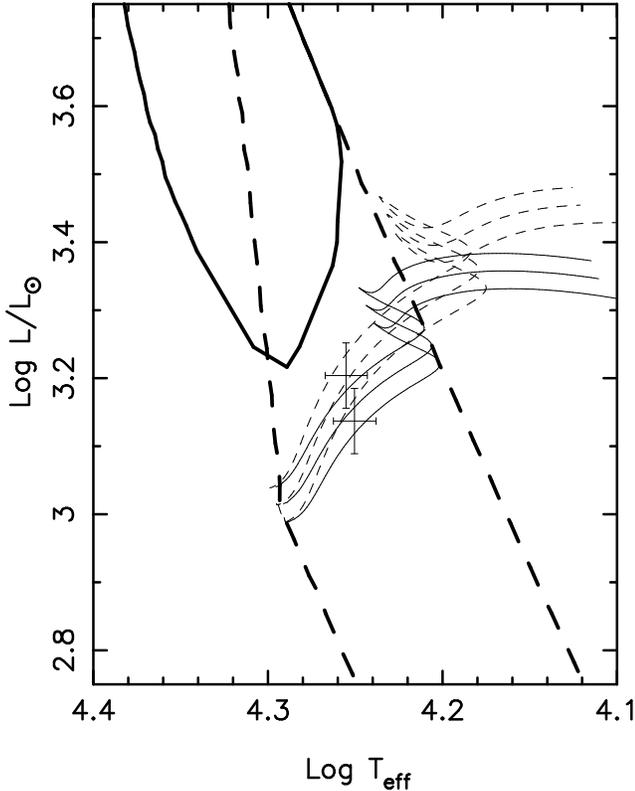}}}
\caption{
The components of CV~Vel in the HR diagram with error bars. The upper point
  represents the primary with $M=6.07\,M_\odot$ and the lower point indicates
  the secondary with $M=5.97\,M_\odot$ (see Table\,\protect\ref{table7}).  The
  full thick line indicates the $\beta\,$Cephei instability strip, the full
  dashed line is the instability strip of the slowly pulsating B stars.  Both
  strips were computed with the non-adiabatic oscillation code MAD (Dupret,
  2001) assuming that no core overshoot occurs.  Evolutionary tracks computed
  with CL\'ES are shown for $M=5.9, 6.0, 6.1\,M_\odot$, and for zero core
  overshoot (thin full lines) as well as a core overshoot of 0.2 (thin dashed
  lines).}
\label{fig4}
\end{figure}

We confronted our observational results for CV~Vel with evolutionary models
computed by De Cat et al.\ (2006) with the Code Lieg\'eois d'\'Evolution
Stellaire (CL\'ES, Scuflaire 2005). These computations were done using the new
solar abundances as reported by Asplund et al.\ (2005) and a standard mixing
length description of convection with $\ell_m=1.75$ times the local
pressure scale
height.  {
The use of these models is appropriate for the components of CV~Vel. Indeed,
the $Z$-values we obtained for the primary and secondary are
$Z=0.0119\pm 0.0028$ and $Z=0.0160\pm 0.0040$ for the case of 
an effective temperature of
18\,000\,K and where we have taken the standard
abundances of Grevesse \& Sauval (1998) for the elements not included in
Table\,6. The metallicities we get are thus consistent with those used in the
model computations by De Cat et al.\ (2006).}
The tracks for $M=5.9, 6.0, 6.1\,M_\odot$ are shown in
Fig.\,\ref{fig4} for two cases: no core overshoot and an overshoot of 0.2 times
the local pressure scale height. We overplotted the position of CV~Vel's
components according to our results mentioned in Table\,\ref{table7}.
{ 
It can be seen that the agreement between our observational 
results and the evolutionary
models is satisfactory.  Recent results from asteroseismology have shown that
a small core overshoot value 
is necessary to bring observed and identified oscillation
frequencies in agreement with evolutionary models (Aerts et al.\ 2003;
Pamyatnykh et al.\ 2004) in this part of the HR diagram. The systematic
uncertainty for the temperatures (and thus luminosities) prevented us from
constraining the overshoot parameters for the components of CV~Vel.  We derive
an age of about 40 million years for the system. Clausen \& Gr\o nbech
(1977) reported an age of only about 30 million years, but they used of course
much older models with somewhat different input physics (e.g., older 
opacity values).  }

\section{Study of the line-profile variability}

The components of CV~Vel reside in the instability strip of the slowly pulsating
B stars, not far away from the $\beta$\,Cephei strip (Fig.\,\ref{fig4}). It is
therefore worthwhile to search for short-period spectroscopic variations of the
components. Percy \& Au-Young (2000) reported the possibility of a variation
with an amplitude of 0.02 mag and a period of 0.25 day in the Hipparcos
photometry. However, they did not study CV~Vel in detail. Clausen (private
communication) was unable to find short-period variations in his extensive
photometric data.

\begin{figure}
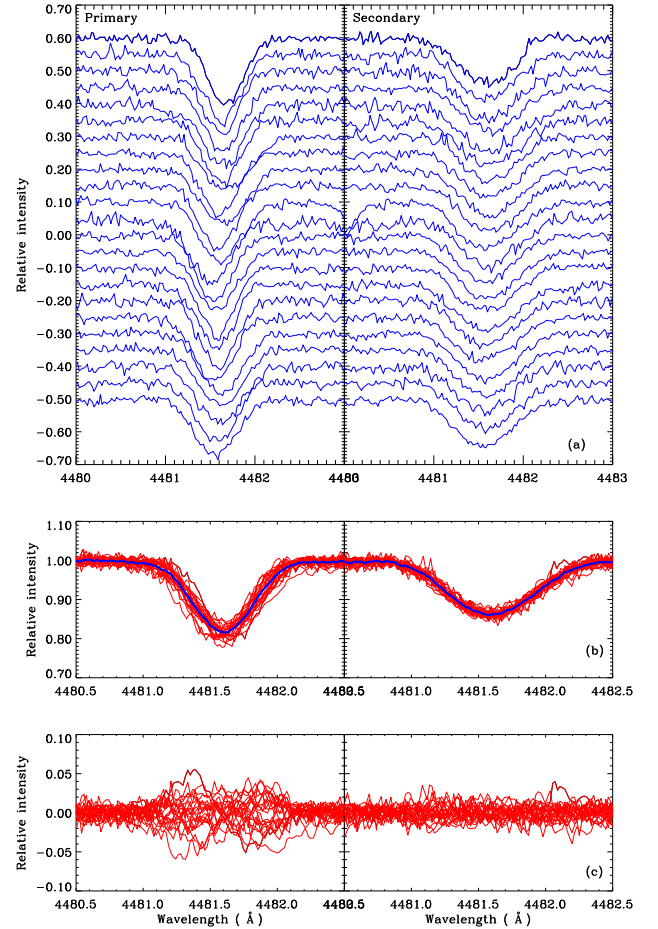

\centering
\rotatebox{0}{\resizebox{8.8cm}{!}{\includegraphics{5506.f5a}}}\\[-0.5cm]
\rotatebox{0}{\resizebox{8.8cm}{!}{\includegraphics{5506.f5b}}}\\[-0.5cm]
\rotatebox{0}{\resizebox{8.8cm}{!}{\includegraphics{5506.f5c}}}
\caption{(a) Normalized Mg II($\lambda$4481) line profiles of the primary (left
panel) and the secondary (right panel).  Phase runs from the top to the
bottom. (b) Line profiles superimposed. The full thick line is the average
profile.  (c) Residual profiles with respect to the average}\label{fig5}
\end{figure}

In Fig.\,\ref{fig5}a we plot the Mg II 4481\AA\ lines of the components
obtained at the phases outside of the eclipses. This line is the
best one to study line-profile variability because it is one of the
deeper ones and has the best S/N level. We shifted all the lines to
the common center of mass. The bold continuous line in Fig.\,\ref{fig5}b is
the average profile. In Fig.\,\ref{fig5}c, the residuals of each line with
respect to the average are shown.  It is apparent from this figure
that the primary shows line-profile variations.

We first of all estimated the rotational and thermal broadening from the average
profiles shown in Fig.\,\ref{fig5}b. It is evident from this figure, as well as
from Fig.\,\ref{fig1}, that the secondary rotates
faster than the primary
since both components
essentially have the same thermal broadening. This is in contrast to Andersen's
(1975) conclusion that the components have equal rotation velocity.
We computed theoretical Mg\,II profiles for Gaussian
intrinsic broadening as well as rotational broadening and find $v\sin i= 19\pm
1$\,km\,s$^{-1}$ for the primary and 31$\pm$2\,km\,s$^{-1}$ for the
secondary. Assuming the orbital and rotational axes to be aligned, and using the
radius estimates from Table\,6, we find a rotation period of 11.0 ($\pm 0.6$)
days for the primary and of 6.4 ($\pm 0.4$) days for the secondary. 
{ It is noteworthy that the primary rotates subsynchronously, while the
secondary's rotation period is compatible with the orbital period within the
errors. Tidal theory (e.g., Zahn 1975, 1977) predicts the occurrence of
synchronisation before circularisation, and both to happen on time scales
shorter than about half the main-sequence duration, depending on the birth
values of the eccentricity and orbital period. Deviations from synchronisation
in circularised binaries have been reported before for close binaries with a
B-type component and a similar orbital period, e.g.\ in the binary $\eta\,$Ori
with a pulsating component (De Mey et al.\ 1996).
}

We tried to quantify the line-profile variability of both components by
computing the line moment variations in the definition of Aerts et al.\ (1992)
for the 23 spectra shown in Fig.\,\ref{fig5}a. These line diagnostics stand for
the centroid variation ($<v>$), the variation in the line width ($<v^2>$) and
the variation in the line skewness ($<v^3>$). They are specifically powerful to
unravel low-degree ($\ell\leq 4$) non-radial oscillation modes. We performed a
frequency search with the Scargle method (Scargle 1982) on $<v>$ in the range
$[0.0,3.4]$\,c\,d$^{-1}$, where the upper limit of this interval corresponds to
the Nyquist frequency. As a result, our dataset is not suitable to detect p-mode
oscillations in B2.5V stars because those have frequencies of typically 4 to
8\,c\,d$^{-1}$ for slow rotators (e.g.\ Aerts \& De Cat 2003). We should be able
to find a signature of g-modes with amplitudes of a few km\,s$^{-1}$ as in
slowly pulsating B stars (e.g.\ De Cat \& Aerts 2002).

The highest frequency peak in the Scargle periodogram of $<v>$ for the primary
occurs at $f = 0.166\pm$0.009\,c\,d$^{-1}$, with an amplitude of
2.2$\pm$0.4\,km\,s$^{-1}$. This frequency has a variance reduction of 54\%,
which implies a decrease in the standard deviation from 2.3 to
1.5\,km\,s$^{-1}$. It reaches a level of 3.8\,S/N, where the S/N level was
computed from an oversampled Scargle periodogram of the residuals over the
entire interval $[0.0,3.4]$\,c\,d$^{-1}$.  We recover the same frequency $f$ in
$<v^3>$. We do not, however, find any periodic phenomenon in $<v^2>$. A
pulsation mode would imply the occurrence of $f$ and/or $2f$ in $<v^2>$ as well
(Aerts et al.\ 1992, De Cat et al.\ 2005). It may be that our dataset is too
limited in quality to detect a frequency in $<v^2>$ given that the even moments
are always noisier than the odd ones (Aerts et al.\ 1992).
\begin{figure}
\centering
\rotatebox{-90}{\resizebox{6.5cm}{!}{\includegraphics{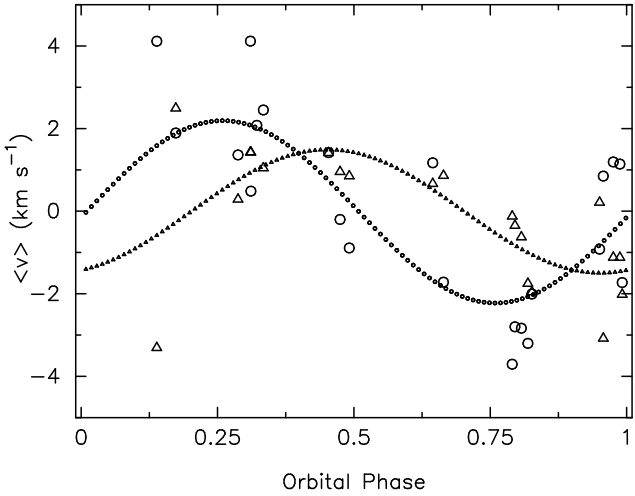}}}
\caption{The first moment variation as a function of orbital phase for the
  primary ($\circ$) and secondary ($\triangle$). The full lines with
  corresponding symbols are a least-squares sine fit.}\label{fig6}
\end{figure}
\begin{table*}
\caption{Observed parameters for B-type main-sequence double-lined eclipsing
binary stars.}
\begin{tabular}{llllllllllll}
\hline
Star            &   Sp.T            &  $P$      &   $M_1$   &  $M_2$    &   $R_1$   &   $R_2$   &$\log T_{e1}$&  $\log T_{e2}$ &   $\log L_1$ &  $\log L_2$ &  Ref \\
\hline
AH Cep          &   B0.5Vn+B0.5Vn   &   1.78    &   15.4    &   13.6    &   6.38    &   5.86    &   4.476   &   4.456   &   4.465   &   4.314   &   1   \\
CW Cep          &   B0.5V+B0.5V     &   2.73    &   13.484  &   12.05   &   5.685   &   5.177   &   4.452   &   4.442   &   4.27    &   4.15    &   2   \\
QX Car          &   B2V+B2V         &   4.48    &   9.245   &   8.46    &   4.289   &   4.051   &   4.377   &   4.354   &   3.72    &   3.58    &   2   \\
V497 Cep        &   B3V+B3V         &   1.20    &   6.89    &   5.39    &   3.69    &   2.92    &   4.290   &   4.249   &   3.265   &   2.860   &   3   \\
V539 Ara        &   B3V+B4V         &   3.17    &   6.239   &   5.313   &   4.432   &   3.734   &   4.26    &   4.23    &   3.29    &   3.02    &   2   \\
CV~Vel          &   B2.5V+B2.5V     &   6.89    &   6.066   &   5.972   &   4.126   &   3.908   &   4.255   &   4.250   &   3.20   &   3.14   &   4   \\
AG Per          &   B4V+B5V         &   2.03    &   5.36    &   4.9     &   2.99    &   4.297   &   4.260   &   4.241   &   2.95    &   2.75    &   5   \\
U Oph           &   B5V+B5V         &   1.68    &   5.186   &   4.672   &   3.438   &   3.005   &   4.22    &   4.199   &   2.91    &   2.7     &   2   \\
DI Her          &   B4V+B5V         &   10.55   &   5.173   &   4.523   &   2.68    &   2.477   &   4.23    &   4.179   &   2.73    &   2.46    &   2   \\
V760 Sco        &   B4V+B4V         &   1.73    &   4.968   &   4.609   &   3.013   &   2.64    &   4.228   &   4.21    &   2.82    &   2.63    &   2   \\
GG Lup          &   B7V+B9V         &   1.85    &   4.155   &   2.532   &   2.411   &   1.753   &   4.176   &   4.041   &   2.42    &   1.61    &   2   \\
$\zeta$ Phe     &   B6V+B8V         &   1.67    &   3.92    &   2.545   &   2.851   &   1.853   &   4.163   &   4.076   &   2.51    &   1.79    &   2   \\
$\chi^2$ Hya    &   B8V+B8V         &   2.27    &   3.605   &   2.631   &   4.384   &   2.165   &   4.066   &   4.041   &   2.5     &   1.79    &   2   \\
IQ Per          &   B8V+B6V         &   1.74    &   3.513   &   1.732   &   2.446   &   1.503   &   4.09    &   3.885   &   2.09    &   0.85    &   2   \\
PV Cas          &   B9.5V+B9.5V     &   1.75    &   2.82    &   2.761   &   2.243   &   2.287   &   4.000   &   4.000   &   1.65    &   1.67    &   2   \\
V451 Oph        &   B9V+A0V         &   2.20    &   2.769   &   2.36    &   2.64    &   2.03    &   4.033   &   3.991   &   1.93    &   1.53    &   6   \\
GG Ori          &   B9.5V+B9.5V     &   6.63    &   2.342   &   2.338   &   1.852   &   1.830   &   3.9978  &   3.9978  &   1.480   &   1.470   &   7   \\
\hline
\end{tabular}
\label{final}\\[0.1cm] {References for Table 7.}: 1 Holmgren et al.\ (1990);
{ 2 Andersen (1991)} and references therein ; 3 Yakut et al.\
(2003); 4 This study; 5 Gimenez \& Clausen (1994); 6 Clausen et al.\ (1986); 7 Torres et al.\ (2000) \\
\end{table*}

For the secondary, we find $f'=1.168\pm$0.009\,c\,d$^{-1}$. This is a one-day
alias of $f$. Forcing $f$ leads to an amplitude of 1.8$\pm$0.4\,km\,s$^{-1}$ (at
level 4.1\,S/N) and a variance reduction of 51\%, bringing the standard
deviation from 1.5 to 0.9\,km\,s$^{-1}$. This time, we cannot recover any
frequencies in $<v^2>$ or $<v^3>$. We show in Fig.\,\ref{fig6} the centroid
velocity $<v>$ for both the primary and secondary, folded with the frequency
$f$, as well as their least-squares fit. We find that the primary's $<v>$ lags
behind a quarter of a phase compared to the one of the secondary.

We conclude to have found evidence of periodic line-profile variability in the
primary and secondary of CV~Vel with a period between 5.7 and 6.4 days. This is
almost equal to the orbital period. Moreover, the quoted frequency error is a
formal 1$\sigma$ least-squares error, and is an underestimation of the true
error given that we covered less than two full cycles with unblended Mg\,II
lines. It therefore seems that line-profile variability occurs with a period
very close or equal to the orbital one.  This is potentially interesting as it
could point towards the excitation of a tidally induced g-mode oscillation in
the components. However, given the phase relation we obtained, and the absence
of any periodic signature in $<v^2>$, it is more likely that a reflection and/or
variable limb-darkening effect lies at the origin of the detected variability.
We need a more extensive data set to firmly establish the correct interpretation
of CV~Vel's line-profile variability.

\section{Summary}

{ 

From combined existing light and old and new radial-velocity curves, we have
obtained a full solution of the double-lined eclipsing binary CV~Vel.  The
system's center-of-mass velocity ($V_\mathrm{0}$) obtained by Feast (1954),
Andersen (1975) and in this study are 27.9(1.1), 24.3(5), and
21.7(2)\,km\,s$^{-1}$, respectively. These three different and decreasing values
may indicate the presence of a distant third body orbiting the binary system.
Forthcoming observations of the system are necessary to evaluate this
hypothesis. We checked but did not find a significant orbital period
change.

We have corrected for the offset in $V_\mathrm{0}$ in Andersen's (1975) and our
data and merged these two high-quality radial-velocity curves to improve our
knowledge on the fundamental parameters of the components. In order to achieve
this, we complemented the spectroscopic data with the high-quality 4-colour
($uvby$) lightcurves obtained by Clausen \& Gr{\o}nbech (1977). These
photometric and radial-velocity data were then solved simultaneously.
The eclipsing and detached properties of the system have led to reliable orbital
and physical parameters in agreement with the values obtained earlier by Clausen
\& Gr{\o}nbech (1977). 

We presented for the first time an abundance analysis for the system and found
results in agreement with those for B stars in the solar neighbourhood for both
components.  This analysis led to a systematic uncertainty of
some 500\,K for the effective temperatures
of the components. Thus, these two parameters of this seemingly well-known
system remain uncertain despite our detailed analysis based on
\'echelle spectroscopy. It is not uncommon to find systematic uncertainties and 
deviations between
photometrically and spectroscopically derived effective temperatures of B-type
stars (e.g.\ De Ridder et al.\ 2004 for a discussion), and even to have
uncertainties of order 500\,K for such stars among methods based on the same
data (e.g.\ Smalley \& Dworetsky 1995, Morel et al.\ 2006, Kaiser 2006). 
Previous works on CV~Vel did not include a spectroscopic temperature
estimate and were thus not able to estimate the 
systematic uncertainty of the component's temperatures. 
}
 
Finally, we discovered line-profile variability in both components. The
variability is most prominent in the primary. From the characteristics of the
line profile moments, we tentatively interpreted this variability as due to an
extrinsic orbital effect rather than an intrinsic oscillation.  Any final
interpretation of the detected line-profile variability requires a much more
extensive dataset, however.

Our results show that CV~Vel consists of young stars which are about half-way in
their central hydrogen burning phase.  Though the orbit is circular, the
components clearly have different rotation speeds. We thus conclude that the
tidal forces had insufficient time so far to bring the system to a circular
orbit with co-rotating components, i.e.\ the circularisation process is
completed but not the synchronisation one.

Observational tests of stellar evolution models have been done by many authors
previously (Pols et al.\ 1997, Young et al.\ 2001, Lastennet \& Valls-Gabaus
2002, Young \& Arnett 2005, and references therein). To determine the masses and
radii of stars independently from models, one should use detached, eclipsing
double-lined binaries and {
compute a simultaneous solution to their radial-velocity and multi-colour
curves.  Unfortunately, accurate physical parameters are only available for very
few systems with early-type components.  Andersen (1991) compiled a list of
stars whose physical parameters are well defined, with relative mass and radius
uncertainties less than 2\%.  We collected all the well-known detached eclipsing
binaries with main-sequence B-type components since then from the literature
(Table\,\ref{final}) and added our new estimates of CV~Vel.
While we obtain a mass estimate with a relative precision of 1.2\%, the
systematic uncertainty for the effective temperature that we derived 
from our
high-quality spectra implies an uncertainty of some 12\% on the luminosities of
the components. Nevertheless, the values we obtained for CV~Vel are fully
compatible with evolutionary models and with those of similar systems whose
parameters are known with better precision.
}
\begin{figure}
\centering
{\resizebox{8.8cm}{!}{\includegraphics{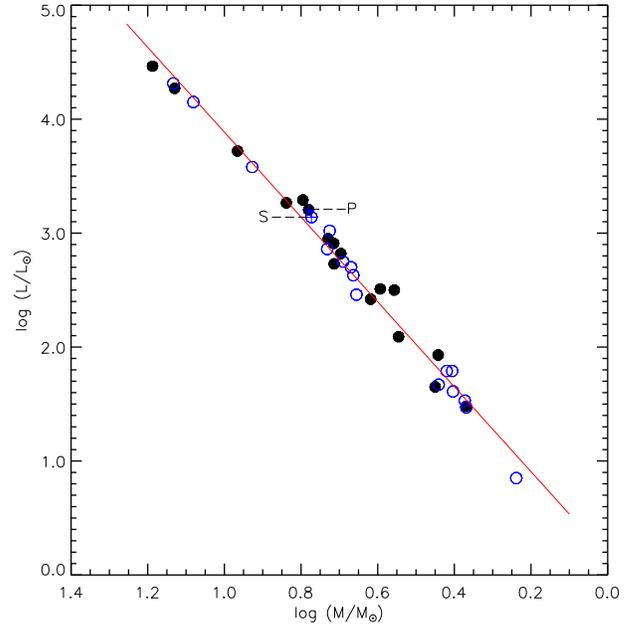}}}
\caption{Mass-Luminosity relation for well-known B-type eclipsing binaries.  The
filled and open circles represent the primary and the secondary components
respectively. The components of CV~Vel are indicated as P and S.}
\label{fig7}
\end{figure}

The stars given in Table\,\ref{final} are drawn in a mass-luminosity plot in
Fig.\,\ref{fig7}. From Table\,\ref{final} we find the following update of the
mass-luminosity relation for the main-sequence B-type stars:
\begin{equation}
\log (L/L_{\odot}) = 3.724(80)\log (M/M_{\odot})+0.162(57).
\label{MLrelation}
\end{equation}
Previously, Hilditch \& Bell (1987) presented a similar relation for O and B
systems. The rms error in log(L) they found was 0.17 whereas it is only 0.11 for
our Eq.\,(\ref{MLrelation}) { because we consider only binaries with B star
components here. This equation is useful to predict masses from luminosities, or
vice versa, for main-sequence B stars.}

\begin{acknowledgements}
The authors are very much indebted to Prof.\ J.\ V.\ Clausen for sharing his
photometric data of CV~Vel with us, for encouraging us to study CV~Vel
spectroscopically and to perform an abundance analysis from our new data. Dr.\
B.\ Vandenbussche is acknowledged for performing the observations, Dr.\ M.\
Briquet for the help in the data preparation and Dr.\ P.\ De Cat for putting his
grid of seismic models at our disposal.  kY would like to thank warmly Drs S.\
Hony, M.\ Reyniers, H.\ Van Winckel and B.\ Kalomeni for their support. The
authors are supported by the Research Council of the University of Leuven under
grant and GOA/2003/04 and a DB fellowship. We are very grateful to an anonymous
referee for numerous comments and helpful constructive remarks which helped us
to improve the paper.
\end{acknowledgements}


\begin{thebibliography}{}
\bibitem[1992]{aerts92} Aerts, C., De Pauw, M., Waelkens, C., 1992, A\&A, 266,
294

\bibitem[2003]{aerts03a} Aerts C., Thoul A., Daszy\'nska J., Scuflaire R.,
Waelkens C., Dupret M.A., Niemczura E., Noels A., 2003, Sci, 300, 1926

\bibitem[2003]{aerts03b} Aerts, C., De Cat, P., Space Science Reviews, 105, 453

\bibitem[1975]{andersen75} Andersen, J., 1975, A\&A, 44, 355

\bibitem[1991]{andersen93} Andersen, J., 1991, A\&ARv, 3, 91

\bibitem[2005]{asplund05} Asplund, M., Grevesse, N., Sauval, A. J., ASP
Conference Series, Vol.\ 336, 25

\bibitem[1994]{balona} Balona, L. A.\ 1994, MNRAS, 268, 119

\bibitem[1996]{baranne} Baranne, A., Queloz, D., Mayor, M., Adrianzyk, G.,
Knispel, G., Kohler, D., Lacroix, D., Meunier, J.-P., Rimbaud, G.,
Vin, A.\ 1996, A\&AS, 119, 373

\bibitem[2007]{briquet} Briquet, M., \& Morel, T. 2007, CoAst,  in press

\bibitem[1977]{cg77} Clausen, J.V., Gr{\o}nbech, B.\ 1977, A\&A, 58, 131

\bibitem[1986]{cg77} Clausen, J.V., Gimenez, A., Scarfe, C., 1986, A\&A, 167,
287
{
\bibitem[2004]{daflon_cunha} Daflon, S., \& Cunha, K. 2004, \apj, 617, 1115
}

\bibitem[2002]{decat02} De Cat, P., Aerts, C., 2002, A\&A, 393, 965

\bibitem[]{} De Cat, P., Briquet, M., Aerts, C., et al.\ 2007, A\&A, 463, 243


\bibitem[]{}
De Cat, P., Telting, J., Aerts, C., Mathias, P.\ 2000, A\&A, 
359, 539

\bibitem[2005]{decat05} De Cat, P., Briquet, M., Daszynska-Daszkiewicz, J.,
Dupret, M.  A., De Ridder, J., Scuflaire, R., Aerts, C.,A\&A, 432, 1013


{
\bibitem[]{}
De Mey, K., Aerts, C., Waelkens, C., Van Winckel, H.\ 1996, A\&A, 310, 164

\bibitem[]{} De Ridder, J., Telting, J.H., Balona, L.A., et al.\ 2004, MNRAS,
351, 324

}


\bibitem[Diaz-Cordoves \& Gimenez(1992)]{1992A&A...259..227D}
Diaz-Cordoves, J., \& Gimenez, A.\ 1992, \aap, 259, 227

\bibitem[]{} Dupret, M.-A., 2001, A\&A, 366, 166

\bibitem[1954]{fm54} Feast, M. W., 1954, MNRAS, 114, 246

\bibitem[1954]{gs55} Gaposchkin, S., 1955, MNRAS, 115, 391

\bibitem[1994]{gimenez} Gimenez, A., Clausen, J. V., 1994, A\&A, 291, 795

{
\bibitem[]{} Grevesse, N., Sauval, J., 1998, SSRv, 85, 161}

\bibitem[1987]{gb87} Hilditch, R. W., Bell, S. A., 1987, MNRAS, 229, 529

\bibitem[1990]{holmgren} Holmgren, D. E., Hill, G., Fisher, W., 1990, A\&A, 236,
409

\bibitem[2005]{hubscher05} Hubscher, J.\ 2005, Informational Bulletin on
Variable Stars, 5643

{

\bibitem[]{}
Kaiser, A., 2006, ASPC, 349, 257

\bibitem[1993]{kurucz93} Kurucz, R. L. 1993, ATLAS9 Stellar Atmosphere Programs
and 2 km\,s$^{-1}$ grid.~Kurucz CD-ROM No.~13.~ Cambridge, Mass.: Smithsonian
Astrophysical Observatory, 1993, 13
}

\bibitem[2002]{lastennet} Lastennet, E., Valls-Gabaud, D., 2002,
A\&A, 396, 551

\bibitem[1985]{leungetal85} Leung, K.-C., Zhai, D., \& Zhang, Y.\ 1985, \aj, 90,
515

{
\bibitem[]{}
Mathias, C., Aerts, C., De Pauw, M., Gillet, D., Waelkens, C.,\ 1994, A\&A,
283, 813

\bibitem[]{}
Mathias, P., Gillet, D., 1993\ A\&A, 278, 511


\bibitem[2006]{morel} Morel, T., Butler, K., Aerts, C., Neiner, C., \& Briquet,
M. 2006, \aap, 457, 651
}

\bibitem[2004]{pam04} Pamyatnykh A. A., Handler G., Dziembowski W. A., 2004,
MNRAS, 350, 1022

\bibitem[2000]{percy} Percy, J. R. \& Au-Yong, K., 2000, Informational Bulletin
on Variable Stars, 4825

\bibitem[1997]{pols} Pols, O. R., Tout, C. A., Schroder, Klaus-Peter,
Eggleton, P. P., Manners, J., 1997, MNRAS, 289, 869


\bibitem[1985]{savage} Savage, B. D., Massa, D., Meade, M., Wesselius, P. R.,
 1985, ApJS, 59, 397

\bibitem[1981]{scargle81} Scargle J. D., 1981, ApJS, 45, 1

\bibitem[2005]{scu05} Scuflaire, R., 2005, CL\'ES User Manual, Version 18,
  Li\`ege University, Belgium

{

\bibitem[]{}
Smalley, B., \& Dworetsky, M. M. 1995, \aap, 293, 446

\bibitem[]{}
Struve, O., Zebergs, V.\ 1960, ApJ, 130, 87


\bibitem[2000]{torres05} Torres, Guillermo, Lacy, Claud H. Sandberg, Claret,
Antonio, Sabby, Jeffrey A., 2000, \aj, 120, 3226

\bibitem[]{}
Van Hoof, A., 1975, PASP, 69, 308
}

\bibitem[1954]{hc50} van Houten, C.J., 1950, Ann. Sterrew. Leiden, 20, 223

\bibitem[1971]{wilson1} Wilson, R. E., \& Devinney, E. J. 1971, ApJ, 166, 605

\bibitem[1994]{wilson94} Wilson R. E. 1994, PASP, 106, 921

\bibitem[2003]{yakut03} Yakut, K., Tarasov, A. E., \.Ibanoglu, C., Harmanec, P.,
Kalomeni, B., Holmgren, D. E., Bozic, H., Eenens, P., 2003, A\&A, 405, 1087

\bibitem[2001]{young01} Young, P. A., Mamajek, E. E., Arnett, D.,
Liebert, J., 2001, ApJ, 556, 230

\bibitem[2005]{young05} Young, P. A. \& Arnett, D., 2005, ApJ, 618,
908

{
\bibitem[]{}
Zahn, J.-P.\ 1975, A\&A, 41, 329

\bibitem[]{}
Zahn, J.-P.\ 1977, A\&A, 57, 383
}


\end{thebibliography}
\end{document}